# Uncovering Synergistic Educational Injustices of COVID-19 and AI

Ahmad Banyasady

Malayer university

## Abstract

Grounded in critical realism and using narrative inquiry, this article explores this article explores the long-term consequences of the COVID-19 pandemic and the rapid proliferation of artificial intelligence within higher education. Through the analysis of student narratives collected in Iranian university settings, the study reveals that learning experiences during and after the pandemic, coupled with unprepared exposure to AI tools, have generated hidden yet impactful layers of educational inequality and cognitive disorientation. These twin phenomena have not only disrupted traditional structures of learning but also carry the potential to deepen existing epistemic and social disparities. The article argues that the university can only fulfill its mission if it develops the epistemic tools necessary to trace layered realities and engage seriously with the often-silenced narratives embedded in students' lived experiences. Such an approach requires a fundamental rethinking in regard to university's role in a rapidly changing educational and technological landscape.

## Introduction

In the wake of the COVID-19 pandemic, higher education systems worldwide underwent rapid—and often disruptive—transformations. While a growing body of international research has examined the immediate impacts of the pandemic—ranging from educational disruption to student well-being (Aristovnik et al., 2020; Crawford et al., 2020; Marinoni et al., 2020, Turnbull, Chugh, & Luck, 2021; Svihus, 2024), less attention has been devoted to its long-term consequences. Recent findings of elevated stress and poor mental health among university students (Bonsaksen et al., 2022; Werner et al., 2021) underscore the importance of devoting greater attention to this vulnerable group. There remains a relative paucity of studies that are context-sensitive or discipline-specific. Issues such as unequal access to online infrastructure across different geographic regions, or the differential efficacy of online learning technologies for students in theoretical fields versus those in practice-intensive disciplines such as engineering and medicine, have not been sufficiently explored (Alkhnbashi et al., 2024). Consequently, it remains unclear what psychological, social, or existential characteristics individuals who lived through the pandemic will carry into the post-COVID era. Accordingly, there is still a lack of clarity regarding the anthropological considerations and policy responses that should be developed to address these complex human dimensions.

These underrepresented context—including universities located in underprivileged areas, lower-tier institutions in national rankings, and systems with highly centralized educational governance—face distinct challenges in navigating the dual pressures of post-pandemic recovery and rapid technological change. For instance, During the COVID-19 period, educational equity was not maintained in under-served areas of Bangladesh. (Roshid et al., 2022). Also, studies indicate that in India, even students enrolled in prestigious universities often struggled to obtain the necessary equipment to access online classes. In several African countries, only 20 percent of universities were able to acquire online infrastructure in response to the COVID-19 pandemic. The availability of such equipment has been linked to deep-rooted and pre-existing structural inequalities, which have, in turn, perpetuated and reinforced educational disparities (Peters et al., 2020). Compared to students enrolled in institutions with greater access to resources and institutional support, those studying at universities outside major centers often experience different learning conditions and receive varying levels of attention in national educational policies and practices. These disparities raise pressing concerns regarding equitable access and inclusion. Access to infrastructure, academic support, and social capital remains highly uneven. Universities benefiting from strong social networks, endowments, and financial flexibility typically enjoy greater institutional autonomy and policy prioritization. Meanwhile, institutions situated in peripheral regions continue to operate under constraints shaped by limited resources, evolving pedagogical demands, and the lingering effects of pandemic-induced educational disruptions.

The second challenge—arguably even more consequential—concerns the rapid integration of artificial intelligence (AI) into teaching, learning, and assessment. While AI is widely promoted as a transformative force promising personalization, scalability, and enhanced pathways for addressing global challenges such as sustainable development (Leal Filho et al., 2023), much of this discourse remains trapped in superficial optimism or binary debates about risks and benefits. Rather than reproducing these narratives, this paper focuses on the less examined socio-educational consequences of AI, particularly in structurally disadvantaged contexts. In under-resourced universities, for instance, AI may be perceived by students as a tool to simulate competence or mask academic insecurity (Long, 2025), potentially complicating the development of authentic academic identity and agency. Although AI is widely regarded as a hallmark of educational innovation, it may, in practice, deepen pre-existing inequalities. This possibility underscores the urgent ethical and institutional responsibility of universities to critically engage with the deployment of such technologies.

This study emerges at the intersection of three intertwined challenges that shape the evolving terrain of higher education. Firstly, the long-term consequences of the COVID-19 pandemic remain underexplored, especially in relation to students whose formative pre-university years had been unfolded under pandemic conditions within structurally under-resourced educational settings. Many of these students now populate classrooms in peripheral universities and bring with them not only unique vulnerabilities but also invisible histories that have yet to be fully understood. Secondly, higher education is entering an era increasingly shaped by artificial intelligence—a force whose implications for learning, identity, and pedagogical agency are still being unfolding. While AI promises personalization and efficiency, it also risks turning into a compensatory mask for students with fragile academic confidence—particularly in under-resourced settings—where it may be used not as a tool of empowerment, but as a means of simulating competence and concealing uncertainty (Long, 2025). Thirdly, and more intimately, as a researcher and educator embedded in a peripheral academic setting, I engage directly with students from various socio-educational backgrounds marked by structural underinvestment. These students have navigated a centralized higher education system, yet their access to educational and social capital remains significantly different from peers in more privileged institutions, including those with greater educational endowments and financial maneuverability.

To illuminate these complexities, I turn to students' narratives of lived narratives not merely as illustrations of experience but as ontological entry points—that is, as experiential traces through which deeper, often invisible structures of inequality and exclusion may be accessed and interrogated. Here, the horizon Bhaskar (2020)[1] outlines for approaching reality is particularly inspiring. In this view, the study does not simply ask how the convergence of post-pandemic legacies and the rise of AI may reproduce or intensify social inequality in higher education; it asks what responsibilities educators and institutions must assume in recognizing and compensating for such disparities—especially within contexts historically pushed to the margins of policy and pedagogical imagination.

---

[1] - Roy Bhaskar presented this paper to the John Templeton Foundation Board Meeting on July 25, 2011, in Edinburgh. It is reproduced here with permission from Roy Bhaskar's literary executors Hilary Wainwright and Mervyn Hartwig. See more: https://www.tandfonline.com/doi/full/10.1080/14767430.2020.1734736#

# Latent Realities in Hidden Narratives:

In this section, I outline the methodological framework guiding this study, which is based on a dual and integrative approach. This integration involves combining narrative inquiry with a critical realist perspective. While such an approach might be viewed as eclectic due to its reliance on multiple frameworks, I offer a reasoned and experience-based justification for its relevance and coherence. Therefore, I think I am on the side of those who consider a plurality of methods and tools essential in conducting research and who speak of critical methodological pluralism and consider it essential for research (Danermark, Ekström & Karlsson, ۲۰۱۹).

The research emerges from a deeply personal and professional concern: as a university instructor, I found the classroom atmosphere and the behavior of the post-pandemic generation of students disorienting and, at times, professionally troubling. If I were to articulate this in more scholarly terms, I would say that I was experiencing a form of disequilibrium—similar to what Piaget describes (Kibler, ۲۰۱۱)—where the observable reality no longer aligned with my existing pedagogical structures. Addressing this sense of cognitive and emotional imbalance became not only a personal necessity but a professional obligation. I found myself at a critical juncture with two divergent paths: to either continue teaching while disregarding the nuanced dynamics within the classroom—thereby reducing the semester to a sequence of midterm and final evaluations—or to adopt a fully intentional pedagogical stance, critically engaging with classroom developments, understanding the diverse profiles of my students, and articulating the breadth of my professional responsibilities. Central to this decision was the imperative to grasp the cognitive frameworks operating among the students—their prior experiences and their current developmental states.

Narrative inquiry, as a qualitative research strategy, provides a meaningful way to engage with these complex experiences. It values the temporal, spatial, and social dimensions of human experience and treats lived realities as valid sources of knowledge (Clandinin, ۲۰۲۲). Rather than focusing on verifiability, this inquiry emphasizes insight, depth, and the capacity of stories to inspire reflection and change (Brookfield, ۲۰۱۷). Nonetheless, one must remain attentive to the potential risks of methodological dominance or interpretive bias (Cizek, ۱۹۹۵). This methodological stance is particularly suited for unpacking the complex interplay between systemic structures and lived experiences—phenomena that cannot be fully accessed through positivist or purely descriptive paradigms. In contexts like post-pandemic Iranian universities, where disruptions have been both material and psychological, such an approach allows for layered interpretation that bridges personal insight and social critique.

The classroom, as the most immediate sphere of my influence, offers a valuable space for exploring how structural and institutional inequalities manifest in students' everyday learning experiences. For example, many of my students come from socially or economically disadvantaged areas in comparison to Iran's central academic hubs. The unequal distribution of digital tools, access to support services, and general academic preparedness reflects a broader pattern of center-periphery imbalance—one that was intensified during the pandemic and remains relevant today. Narrative inquiry allows me to reframe and reflect on my professional identity in this context. It supports a pedagogy rooted in critical awareness, where the educator does not merely transmit knowledge but actively interrogates the power structures and hegemonic assumptions that shape classroom life (Ashwin et al., ۲۰۲۰; Brookfield, ۲۰۱۷). Through this process, the teacher becomes a reflective practitioner, engaging with the deeper social and institutional structures behind surface-level challenges.

This reflective approach is enriched by the ontological framework of critical realism, particularly as formulated by Roy Bhaskar. Critical realism distinguishes among three levels of reality: the empirical (what is experienced), the actual (what occurs, whether observed or not), and the real (the underlying structures and mechanisms that generate events) (Bhaskar & Hartwig, ۲۰۱۰; Shipway, ۲۰۱۰). From this standpoint, knowledge is not confined to perception alone but relates to mechanisms that operate independently of our awareness (Bukowska, ۲۰۲۱). Applying this lens enables researchers to go beyond mere description and uncover causal relationships that shape educational phenomena (Fletcher, ۲۰۱۶; Yeung, ۱۹۹۷), as it could improve our understanding about human life and social structures (Bhaskar, ۲۰۲۰; Bukowska, ۲۰۲۱). Although I do not follow a strict procedural model of critical realist methodology, I remain committed to its foundational logic, that could lead to better theoretical development and research procedure (Easton, ۲۰۱۰). My goal has been to increase conceptual clarity and analytical coherence in a way that highlights contradictions, tensions, and asymmetries within the teaching context (Yamagata-Lynch et al., ۲۰۱۷; Pino Gavidia & Adu, ۲۰۲۲).

Another conceptual lens that informs this study is fuzzy logic—originally developed by Lotfi Zadeh (١٩٨٨)—as an approach that incorporates uncertainty in our reasoning process. Unlike binary logic, fuzzy thinking allows for degrees of truth, offering a more nuanced way to understand complex, ambiguous, and fluid realities (Kosko, ١٩٩٣). While widely applied in engineering, its potential in the humanities remains largely underexplored, despite the high capacity of these fields (Bakhov et al., ٢٠٢١). In the context of this research, fuzzy logic serves not as a technical method but as an epistemological metaphor that legitimizes partiality, ambiguity, and the in-between spaces where marginalized narratives often reside.

Thus, while this methodological approach was developed within a specific and localized university context, I believe its conceptual insights and reflective orientation can illuminate pedagogical praxis in other settings marked by similar social and geographical disparities. In a globalized era shaped by shared educational disruptions, narrative inquiry remains a powerful tool for deepening pedagogical understanding and fostering critically engaged practice.

## Repository of Hidden Narratives

In the broader landscape of contemporary educational discourse—especially those influenced by trends in popular psychology—there is a growing tendency to label students without adequate empirical grounding. A frequent example is the premature diagnosis of children as having attention-deficit disorders, even when their behavior reflects normal developmental energy (Brookfield, ٢٠١٧). When such assumptions go unchallenged, they can become naturalized within institutional routines, narrowing the interpretive horizon of educators and policymakers alike. It was within this epistemic climate that I began to observe a puzzling pattern in my own university classrooms in three semesters after the COVID-١٩ pandemic. Although students regularly affirmed the clarity and dynamism of my teaching strategies—a point I raise not out of self-congratulation but to contextualize their response—they nevertheless exhibited a widespread sense of disengagement. Their interactions with course content were marked by emotional flatness, passivity, and what I came to perceive as a kind of silent fatigue. Over time, this atmosphere of dormancy began to impact my own pedagogical presence.

In response, I revised my lesson design to include short segments on personal development and student well-being. I introduced varied activities at the start of each session in an effort to spark interest and restore classroom energy. Initially, these changes showed promising results. Students left sessions more animated, and many returned in informal contexts—hallways, office hours—to share thoughts or discuss personal concerns. Yet the effect proved transitory. Within weeks, I was once again facing a group of students who appeared emotionally and cognitively distant. At first, I turned to commonly cited explanations within academic and media discourses: these students were members of "Generation Z"—digitally saturated, socially fragmented, and psychologically shaped by virtual spaces (Twenge, ٢٠١٧). While this generational framing offered a convenient lens, it ultimately proved inadequate. It reduced a complex phenomenon to deterministic traits and offered little in the way of pedagogical direction.

If I am to express my experience in more structured psychological terms, I would say that I entered a state resembling what Piaget described as disequilibrium—a moment when the existing cognitive schema fails to accommodate lived reality, thereby prompting the need for reconstruction. Faced with this dissonance, I began paying closer attention to the subtle cues—gestures, language, emotional tonality—that emerged in my classroom. What I encountered was revealing: signs of anxiety, mental exhaustion, and emotional disconnection were recurrent. I suspected there were deeper, unarticulated layers of experience shaping student behavior. Embracing the critical realist perspective (Bhaskar, ٢٠٠٨; Fletcher, ٢٠١٦), I recognized the importance of probing beneath the empirical surface to reach the generative mechanisms behind the observed outcomes. This realization prompted me to initiate informal classroom dialogues. I began asking students about their daily lives—what media they consumed, how they spent their leisure time, what they found meaningful. During one such session, I spontaneously asked about their experience of the COVID-١٩ pandemic. The effect was immediate: the classroom atmosphere transformed, and guarded silence gave way to energetic sharing.

A student said bluntly, "Everything changed after COVID." Another revealed a GPA drop from ١٩٫٧ to ١٧. One recounted how they had been an avid reader until they received a smartphone; since then, they had stopped reading entirely. Others spoke of attending online classes by logging in and going back to sleep, or of memorizing poetry in front of mirrors to satisfy anxious teachers. Students from rural areas described limited access to internet infrastructure and the emotional burden of navigating education through mobile phones—often the only available device in their households. While these stories emerged in informal, unscripted contexts, they reveal patterns that demand critical theoretical engagement. They are not isolated anecdotes, but symptomatic of broader epistemological ruptures—ruptures that challenge dominant pedagogical frameworks and call for a rethinking of how equity, access, and educational meaning are defined in AI-driven, post-pandemic learning environments.

In these exchanges, it became clear that many students' entire educational engagement during the pandemic had taken place via smartphones, not laptops or tablets. These stories echoed structural inequalities already embedded in the geography of Iran's higher education system—where, despite the geographic spread of universities, access to infrastructure and educational opportunities remains uneven. In this regard, the convergence of digital dependence and unequal resource distribution has created a post-pandemic educational terrain fraught with new forms of vulnerability. These stories reveal a crucial yet underexplored gap within contemporary educational research. While the literature has begun to recognize the psychological and technological consequences of the pandemic (Rapanta et al., ۲۰۲۱; Cattaneo Della Volta et al., ۲۰۲٤), it rarely addresses the compounding effects of COVID-۱۹ experiences and the rapid introduction of AI-driven systems on marginalized student populations—particularly in non-Western, resource-asymmetric contexts. My classroom became a site where these tensions surfaced—not in formal surveys or test scores, but in unplanned conversations and shared reflections. What these stories reveal is not merely anecdotal. They signal the presence of deeper systemic ruptures that remain unaddressed in dominant pedagogical models. If we are to create equitable and inclusive AI-integrated learning environments, we must first develop a richer understanding of the lived realities of students who were educationally formed during the pandemic. Ignoring these histories not only obscures students' present realities but also risks perpetuating the very inequities AI-driven educational reforms aim to dismantle.

## Reflective Encounters with the Post-COVID Generation

A review of Iranian research on the COVID-۱۹ pandemic reveals a notable pattern: the majority of studies have primarily concentrated on the period during the outbreak itself. This emphasis is understandable given the emergency-driven and unpredictable nature of the pandemic. However, what stands out is the relative alignment of many Iranian studies with the global body of COVID-۱۹ research; local findings were often documented in a timely manner and with acceptable scientific rigor. For example, a study by Abolmaali Alhosseini (۲۰۲۰), consistent with global research (Brooks et al., ۲۰۲۰), pointed to feelings such as hopelessness, fatigue, and boredom as psychological consequences of the pandemic among school-aged students. Moreover, a number of studies have explored the multifaceted challenges faced by students and their families. These challenges include inadequate infrastructure for virtual education, limited familiarity with digital tools, barriers to effective participation in online classes, and the inability of educational institutions to fully digitize instructional content (Sahebi & Elyasi, ۲۰۲۲).

Furthermore, empirical evidence suggests that even after students returned to in-person learning environments, the psychological effects of the pandemic remained visible in their behavior. These include increased anxiety, decreased social interaction, and diminished communication skills (Javaheri, Vakili, & Esmaily, ۲۰۲۲). Nonetheless, a critical point that deserves attention is the predominance of cross-sectional studies over longitudinal ones in this field. In particular, there is a noticeable scarcity of research that investigates students lived experiences during the transition from pre-university education to higher education within the context of the pandemic. This research gap is noteworthy, especially considering that longitudinal studies can over time reveal deeper and more enduring impacts of the pandemic on the educational, psychological, and social trajectories of vulnerable generations.

## Rapid entry of the second wave

As previously noted, while the long-term—and in some cases, enduring—consequences of the COVID-۱۹ pandemic continue to demand analytical and practical attention, Iranian higher education, like that of many other parts of the world, is now encountering a new wave of technological transformation: the rapid and pervasive emergence of artificial intelligence (AI). This shift is occurring at a time when post-pandemic research—both globally and within Iran—has yet to reach theoretical and practical maturity, particularly in underserved regions where the pandemic's impacts remain insufficiently examined and inadequately addressed. Each of these two phenomena, owing to their distinct characteristics, can produce divergent effects across different social and geographical contexts. For example, in communities where families, due to cultural values, prioritize early exposure to the English language and possess the financial capacity to pursue such goals, stark disparities exist in comparison to families who either lack this orientation or are unable to access the necessary financial or institutional support, such as reputable language institutes. These disparities can progressively and exponentially widen the gap between social groups over time. Similar patterns can be observed in the realm of access to educational technologies, where unequal distribution exacerbates existing socio-economic inequalities. Should the residual effects of the pandemic and the accelerating development of AI unfold concurrently, the combined impact could significantly amplify existing inequities. Consequently, the simultaneous confrontation with the lingering challenges of COVID-۱۹ and the emergent demands of AI underscores the urgent need to rethink and reform the structures, policies, and programs of higher education.

In contrast to the aforementioned global challenges, Iranian higher education also faces a set of chronic, structural issues that merit serious attention. One of the most critical barriers is the restricted access to key AI technologies—a problem largely caused by recent political and military sanctions. As a result, the cost of accessing AI tools in Iran is substantially higher compared to other regions. Furthermore, certain AI platforms have imposed specific restrictions on Iranian users, making their use practically infeasible without circumventing technological filters and regulatory constraints. Adding to this complexity is the structure of Iranian academia, which remains predominantly publication-driven. Academic recognition and institutional ranking are strongly tied to the number of articles published in foreign, particularly English-language, journals. This has created systemic incentives that privilege international publication. At the same time, although English instruction begins at early educational stages in Iran, it has not yielded proportionate linguistic competencies among learners, leading many to rely on costly private institutions that operate outside the official public education system. Within such a context, the integration of AI into Iranian higher education—especially given its algorithmic biases and known challenges—may present significant threats to equity and effectiveness. While these challenges can be studied at the national level, it is crucial to emphasize that their impact is often disproportionately greater in underserved and marginalized regions.

Numerous studies in Iran's higher and general education sectors have identified a range of barriers hindering effective AI adoption. These include insufficient legal, hardware, and software infrastructures; a shortage of qualified human resources; slow internet speeds; internet filtering; algorithmic biases; ethical concerns; and violations of data privacy (Hamedinasab & Rahimi, ۲۰۲۵; Ahmadi et al., ۲۰۲۵; Hoseini Moghadam, ۲۰۲۳; Adel, ۲۰۲٤; Daneshvar Heris, ۲۰۲٤). Other documented obstacles include cultural resistance to change (Hamedinasab & Rahimi, ۲۰۲۵), the lack of strategic policy making to promote data-driven decision-making (Hoseini Moghadam, ۲۰۲۳), and the absence of proactive university engagement through training programs for faculty, students, and administrators (Ahmadi et al., ۲۰۲۵; Daneshvar Heris, ۲۰۲٤). The rapid and largely unregulated use of AI tools by both students and faculty—without proper governance frameworks—poses a serious threat to the credibility of assessment systems, academic integrity, and the overall teaching–learning process.

In response to these concerns, some studies have proposed proactive strategies to manage this technological shift. These include raising awareness, providing targeted training for faculty, students, and administrators, developing specialized academic programs in AI, fostering a culture of data-informed decision-making, reforming institutional structures, creating supportive legal frameworks, and implementing broad reskilling initiatives (Hosseini Moghadam, ۲۰۲۳; Olyaee, et al., ۲۰۲٤). However, implementing such recommendations in under-resourced areas presents additional challenges, such as weak communication infrastructure, limited access to smart devices, and a lack of foundational digital literacy education. These barriers contribute to a renewed manifestation of educational inequality and structural discrimination within higher education.

Thus, the interplay between lingering post-pandemic challenges and the emerging opportunities and risks associated with AI has created a complex, multi-layered reality. A critical analysis of this dual transformation is essential to shaping the future trajectory of higher education.

## Layers of reality & policy making windows

In recent years, the Islamic Republic of Iran has sought to address the multifaceted challenges posed by artificial intelligence (AI) through high-level policy initiatives, most notably the formulation and ratification of the National AI Strategy Document. Characterized by its ambitious and idealistic tone, this document sets forth strategic goals such as positioning Iran among the world's top ten AI powers by the year ۲۰۳۳–۳٤. At the heart of its vision lie objectives including accelerated development, wealth creation, and the enhancement of societal welfare and national security. The document also underscores the role of AI in improving governance quality, with particular emphasis on social justice, national security, sustainability, social cohesion, and the strengthening of social capital (National Artificial Intelligence Document of the Islamic Republic of Iran, ۲۰۲۳).

Further structural commitment is reflected in the Iranian Parliament's subsequent initiative to establish a National AI Organization (Islamic Parliament Research Center, ۲۰۲٤), signaling a formal effort to institutionalize AI governance. Nevertheless, persistent challenges—such as structural poverty, centralized bureaucracy, and regional disparities—continue to impede the effective implementation of AI strategies, particularly in the domains of higher and public education. Reports indicate a decline in household spending on education among lower-income deciles, with poor families facing growing financial constraints that limit their investment in their children's learning. Overcrowded classrooms, teacher shortages, and reduced per-student public funding—particularly amid rising student populations—have negatively affected learning outcomes. In the post-COVID era, the digital divide has become a defining barrier, as limited access to the internet, computers, and tablets severely impacts low-income communities. The pandemic highlighted how unequal access to technology can deepen educational poverty and disrupt learning continuity. As both public education budgets and household expenditures shrink, educational inequality increases: wealthier families can still afford high-quality education, while poorer segments face restricted access and declining quality (Sheikh Razi, ۲۰۲۳).

To mitigate educational poverty and improve learning outcomes for students from low-income households, policy experts (Sheikh Razi, ۲۰۲۳) have proposed that the Ministry of Education and the Ministry of Cooperatives, Labor, and Social Welfare implement multifaceted strategies. These include identifying at-risk families, providing conditional financial support, offering integrated social services, and delivering family-focused counseling—particularly aimed at transforming underperforming schools in high-poverty areas into high-achieving institutions.

But These policy efforts have emerged in a context still deeply shaped by the long-term and structural consequences of the COVID-۱۹ pandemic. The crisis exposed and exacerbated pre-existing inequalities in Iran's education system, particularly regarding access to digital tools and infrastructure. In this post-pandemic landscape, realizing the lofty ambitions of national AI policies requires serious engagement with the lingering consequences of COVID-۱۹—especially those related to equity, infrastructure, and educational access.

In summary, while top-down initiatives such as the National AI Strategy and the proposed National AI Organization reflect a bold and aspirational policy vision, the realization of their goals depends heavily on addressing the structural inequities exacerbated by the COVID-۱۹ crisis and rebuilding the educational system through a justice-oriented and resilient approach. In today's world, equitable distribution of resources and opportunities is no longer regarded as the ultimate solution for achieving social justice. Although macro-level policies still emphasize fair allocation, they often face significant practical limitations. It can be firmly asserted that many of these approaches—particularly those framed as symbolic legislation or declarative statements—prove ineffective in implementation.

This issue is especially critical in the realm of AI, where access to smart technologies is not invariably dictated by governmental policies but by individual financial capacity. While smartphones, the internet, and by extension AI may appear universally accessible, this superficial equality does not translate into genuine equity, nor does it meaningfully improve social equality. Worse still, over time, such illusory parity may entrench and exacerbate systemic inequalities, leaving profound negative impacts on societal structures. Therefore, policies must be formulated and refined through a profound and comprehensive understanding of the public sphere and the systemic currents that generate chronic and exponentially compounding inequities. Achieving this objective is only possible through meticulous and thorough recognition of the various layers of reality within the processes of theorization and policymaking. To this end, the functions, interactions, and synergistic effects of these realities must be properly articulated and comprehended so that policies can be effectively and productively designed and implemented.

## The University and Hidden Realities

The COVID-19 pandemic left a vast and multi-layered imprint on human life—an impact that resists reduction to a single, uniform experience. Rather, it unfolded across a wide spectrum: from radical disruptions such as the loss of loved ones, severe illness, and traumatic hospitalizations, to more subtle yet enduring psychological consequences. These include stress related to vaccine access, distrust of health information, and confusion caused by conflicting news. On a deeper level, persistent psychological effects—such as anxiety, fatigue, cognitive exhaustion, declining academic motivation, and detachment from dialogical and communal learning—have reshaped educational trajectories. Such experiences challenge the very assumptions of authentic learning, which many believe can only occur within contexts of active engagement and interpersonal interaction.

Despite these layered disruptions, students and educators were expected to continue the educational process under extraordinary and unnatural conditions—ranging from full school closures to hybrid models of learning and various forms of social distancing. The burden extended beyond students to parents and teachers, many of whom had no prior experience with digital pedagogy. Structural inequalities—such as regional disparities in infrastructure and varying levels of family resources—exacerbated the situation. A large number of families were forced to absorb unexpected financial pressures to provide mobile phones, tablets, or stable internet connections. Importantly, it should be recalled that, prior to the pandemic, the use of mobile devices was either prohibited or heavily restricted in most educational settings. COVID-19 rendered such boundaries porous—if not obsolete—virtually overnight, signaling a profound shift in educational norms.

These inequities manifested with even greater intensity in marginalized or under-resourced areas, where poverty, digital illiteracy, and infrastructural limitations created tangible and structural exclusions. Among the most critical and overlooked challenges was media literacy. In policy frameworks, institutional planning, and even public discourse, digital competence was treated as if it would emerge spontaneously. Yet in practice, many learners were submerged in a sea of unfiltered, unreliable, and often misleading information. Students, now with unprecedented access to platforms like Instagram and with minimal oversight, found themselves adrift—without the necessary training to navigate digital spaces productively. The persuasive designs of these platforms—engineered to retain user attention—stood in stark contrast to the unfamiliar, under-resourced, and pedagogically uninspiring environments of formal schooling.

Repeated Connectivity losses, unstable platforms, and technical interruptions—especially prevalent in underprivileged regions—frequently disrupted learning. Often, just being "in class" became an illusion rather than a guarantee of meaningful engagement. Moreover, mere attendance, even under ideal conditions, is never equivalent to actual learning. Compounding the problem was the absence of clear and universally adopted standards for online education. While public health measures such as social distancing were communicated through widely visible signs and protocols, online education lacked comparable guidance. In a country like Iran, marked by vast geographic and socio-economic diversity, access to a tablet or reliable internet effectively determined whether a student remained within or outside the sphere of education. This digital divide created a new form of educational exclusion—silent, invisible, yet structurally entrenched.

We are now in a post-COVID era, yet we face a generation of university students whose cognitive, social, and emotional development was shaped under conditions of isolation, anxiety, and pedagogical fragmentation. For many, schooling took place not in collaborative or social environments, but in private rooms, in front of screens, in a semi-passive state. These students have now entered a university system that still relies heavily on traditional expectations: physical presence, written assignments, and classroom participation. But these expectations are not always aligned with the educational trajectories formed during the pandemic. Some students have become accustomed to online multiple-choice assessments, peer-sharing of answers via messaging apps, and minimalist study habits. I myself witnessed this during my teaching in a teacher training center, where patterns of identical exam responses led me to discover that answers were being circulated in real time through WhatsApp groups.

These students—products of what might be called "COVID education"—now face a very different academic landscape. Faced with analytical assignments, project-based evaluations, and in-person demands, many of them report heightened anxiety, reduced motivation, and cognitive disorientation. And while they struggle to adjust, their university years are quietly passing—without the formation of a meaningful or authentic academic identity.

Amid this fragile and fragmented context, the abrupt and unregulated influx of artificial intelligence into higher education has added yet another layer of complexity. Today, with just a smartphone and basic internet access—which most university campuses provide—students can access powerful AI tools like ChatGPT. In the absence of structured guidance and critical engagement, many students resort to these technologies as quick fixes. My experience reveals that even at the graduate level, students submit essays that appear polished but rely on fabricated or unverified sources. Without intervention, this trend risks turning into an epistemic crisis—one that erodes academic integrity and weakens the pedagogical core of higher education.

Where, then, do we stand? What do these layered narratives and converging disruptions demand of the university? What responsibility must higher education institutions assume? While encountering these realities may change our perspective, it is far from sufficient. We must come to terms with the fact that we have moved beyond a world of predictability and certainty. This is not a return to earlier, simpler times; rather, we are witnessing an era in which the pace of change is itself a force of disorientation. The collision between a "COVID-injured educational body" and the aggressive infiltration of AI is no longer an abstraction—it is now a tactile reality. Who could have imagined that so soon after the pandemic, educational systems would be so deeply entangled with AI technologies, evolving week by week? The competition between corporations and platforms reveals a foundational transformation. Consider, for instance, how the emergence of models like DeepSeek caused abrupt market shifts among AI-based firms (Clark,۲۰۲۵; Stone, ۲۰۲۵).

As a non-native English speaker with over fifteen years of academic experience in international settings, I have long struggled with the demands of translation and cross-cultural communication. Crafting a simple email used to require hours of searching for the right words. Today, I write my ideas in informal Persian, feed them into a simple AI interface, and receive—in seconds—a coherent, polished academic response that I can confidently send to colleagues across the globe. During the pandemic, I began participating in global online seminars on different platforms like zoom, often worried whether I truly understood native English speakers or whether my questions would be clear. Today, those anxieties have faded. I attend sessions regularly, draft my thoughts quickly, and receive meaningful feedback—feedback that confirms not only comprehension but connection.

All of this signals a profound transformation. But is this where change stops? Certainly not. It is only the beginning. The shifts we are witnessing are not only irreversible; they are accelerating, deeply interwoven, and demand that we reimagine the future of education in radically new terms. At this point, we must ask: what is the mission of the university—and of the academic—in such a context? To respond, I once again draw upon the conceptual tools offered by critical realism, especially Roy Bhaskar's notion of a "depth ontology" and the need for reality-tracing instruments.

Here, a conceptual bridge can be drawn between Bhaskar's stratified ontology of reality and Zadeh's (۱۹۸۸) fuzzy logic framework—two distinct yet complementary approaches that help us move beyond binary, surface-level interpretations. While Bhaskar invites us to consider the layered depths of the real, Zadeh's work encourages recognition of the ambiguous, fluid boundaries within which contemporary realities operate. The reality we now confront is multilayered. Its surface layers can deceptively sustain the illusion of a familiar higher education system. Indeed, many universities, post-COVID, swiftly returned to pre-pandemic modes of instruction without fully recognizing the hidden transformations brought about by students disrupted learning experiences or the overwhelming infiltration of artificial intelligence.

When we add to this reality the ongoing uncertainty, the deepening complexity, and the accelerating pace of transformation, it becomes evident that universities must equip themselves with the cognitive and structural capacity to perceive and respond to deeper layers of change. The academic of the present and future cannot remain a passive observer. Rather, they must act as active agents—ready to face not only the unknown but also the potentially destabilizing events that may lie ahead. Whether it is the intensification of geopolitical conflicts in Ukraine, Palestine, or elsewhere, or the emergence of new disruptive uses of AI by states and corporations, the message is clear: surprise can no longer be part of the university's operational vocabulary.

The vitality of the university can only be sustained if it becomes epistemologically and institutionally capable of engaging with the hidden and advancing dimensions of reality. Leveraging the conceptual apparatus of critical realism—especially the distinction between the empirical, the actual, and the real—may offer a meaningful starting point for reimagining the role of the university in a world that is not only unpredictable, but fundamentally transformative.

# Conclusion

This article has demonstrated that the long-term and often underexplored consequences of the COVID-19 pandemic, coupled with the rapid and widespread emergence of AI in education, have created a complex, multilayered, and increasingly unstable landscape for higher education systems. These complexities invite a profound rethinking of how we perceive and interpret reality—one that requires theoretical frameworks capable of engaging with deeper ontological layers, such as critical realism. At present, we do not fully understand the enduring cognitive, psychological, and behavioral traits that post-COVID students may carry. Nor do we know to what extent students' engagement with AI technologies reflects efforts to enhance their learning capacities, or rather attempts to compensate for underlying deficits—such as diminished self-confidence—rooted in broader social inequalities. universities must play a more active role in fostering epistemic insight, amplifying hidden narratives, and redefining themselves as dynamic and responsive institutions capable of navigating complex human futures.

While this article attempted to envision the mission of the university through the lens of classroom-based narratives, it is important to acknowledge a key limitation: the data were collected from a limited number of students within a specific geographic and institutional context, over the course of three academic terms. This naturally restricts the scope and generalizability of the findings. Future research—both qualitative and quantitative—is therefore essential to deepen our understanding of these emerging realities. Such studies will help develop a more comprehensive picture of the evolving dynamics of post-pandemic higher education, particularly in relation to the ethical, psychological, and institutional implications of artificial intelligence.